\documentclass[12pt]{article}
\usepackage{amsmath}
\usepackage{amsmath,bm}
\usepackage{url}
\usepackage{tikz}
\usetikzlibrary{shapes.geometric}
\usetikzlibrary{arrows}
\usepackage{amsmath}
\usepackage{amssymb}
\usepackage{hyperref}
\usepackage{rotating}
\usepackage{physics}
\usepackage{epsfig}
\usepackage{graphicx}
\usepackage{color}
\usepackage{cite}
\usepackage[font=small,labelfont=bf]{caption}
\usepackage{xcolor}

\newcommand{\beq}{\begin{equation}}
\newcommand{\eeq}{\end{equation}}
\newcommand{\ga}{\lower.7ex\hbox{$\;\stackrel{\textstyle>}{\sim}\;$}}
\newcommand{\la}{\lower.7ex\hbox{$\;\stackrel{\textstyle<}{\sim}\;$}}
\newcommand{\tabitem}{~~\llap{\textbullet}~~}

\hypersetup{
    colorlinks = true,
    citecolor = {blue},
    linkcolor = {blue},
    urlcolor = {blue},
}

\begin{document}

\def\jcap{\ref@jnl{J. Cosmology Astropart. Phys.}}

\begin{flushright}
{\tt KCL-PH-TH/2019-51}, {\tt CERN-TH-2019-089}  \\
{\tt ACT-03-19, MI-TH-1923} \\
{\tt UMN-TH-3826/19, FTPI-MINN-19/17} \\
\end{flushright}

\vspace{0.2cm}
\begin{center}
{\bf {Unified No-Scale Attractors}}

\end{center}
\vspace{0.1cm}

\begin{center}{
{\bf John~Ellis}$^{a}$,
{\bf Dimitri~V.~Nanopoulos}$^{b}$,
{\bf Keith~A.~Olive}$^{c}$ and
{\bf Sarunas~Verner}$^{c}$}
\end{center}

\begin{center}
{\em $^a$Theoretical Particle Physics and Cosmology Group, Department of
  Physics, King's~College~London, London WC2R 2LS, United Kingdom;\\
Theoretical Physics Department, CERN, CH-1211 Geneva 23,
  Switzerland;\\ 
  NICPB, R\"avala pst. 10, 10143 Tallinn, Estonia}\\[0.2cm]
{\em $^b$George P. and Cynthia W. Mitchell Institute for Fundamental
 Physics and Astronomy, Texas A\&M University, College Station, TX
 77843, USA;\\ 
 Astroparticle Physics Group, Houston Advanced Research Center (HARC),
 \\ Mitchell Campus, Woodlands, TX 77381, USA;\\ 
Academy of Athens, Division of Natural Sciences,
Athens 10679, Greece}\\[0.2cm]
{\em $^c$William I. Fine Theoretical Physics Institute, School of
 Physics and Astronomy, University of Minnesota, Minneapolis, MN 55455,
 USA}
 
 \end{center}

\vspace{0.1cm}
\centerline{\bf ABSTRACT}
\vspace{0.1cm}
{\small We have presented previously a general treatment
of Starobinsky-like inflation in no-scale supergravity where the tensor-to-scalar ratio $r = 3(1 - n_s)^2$, and $n_s$ is the tilt of the scalar perturbations.
In particular, we have shown how this scenario can be unified with modulus fixing, supersymmetry breaking and a small cosmological constant.
In this paper we extend these constructions to inflationary models based on generalized no-scale structures. In particular, we consider
alternative values of the curvature parameter, $\alpha < 1$, as may occur if not all the complex K\"ahler
moduli contribute to driving inflation, as well as  $\alpha > 1$, as may occur if complex structure moduli also contribute to driving inflation.
In all cases, we combine these $\alpha$-Starobinsky inflation models with supersymmetry breaking and a present-day
cosmological constant, allowing for additional contributions to the vacuum energy from stages of gauge symmetry breaking.
 } 

\vspace{0.2in}

\begin{flushleft}
June 2019
\end{flushleft}
\medskip
\noindent

\newpage

\section{Introduction}

Inflation \cite{reviews} provides a very successful phenomenological framework for describing the large-scale structure of the universe,
including its near-flatness, the smallness of the primordial density perturbations measured in the cosmic microwave
background (CMB), their small tilt $n_s \simeq 0.965$ and
their consistency with a Gaussian white-noise spectrum \cite{planck18}. That said, further tests remain to be made, e.g., via the
existence and spectrum of tensor perturbations \cite{rlimit}. However, the successes of inflation motivate attempts to relate it to
the Standard Model (SM) of laboratory particle physics on the one hand and, on the other hand, to a candidate quantum theory of 
everything including gravity, such as string theory. The characteristic energy scale of inflation is presumably intermediate between 
those of the SM and quantum gravity, and models of inflation may provide a welcome bridge between them.

These considerations have long since been a motivation for models of inflation  \cite{GL,KQ,EENOS,otherns} based on no-scale supergravity
\cite{no-scale,EKN1,EKN2,LN}.
This is the generic form of effective field theory that arises in the low-scale limit of string compactifications \cite{Witten},
and the persistence of supersymmetry down to a scale below the inflation scale would help its stabilization \cite{cries}, as
well as that of the electroweak scale \cite{natural}.

The upper limit on the ratio $r$ of tensor perturbations in the CMB relative to the measured scalar density perturbations \cite{rlimit} is entirely
consistent with the original model of inflation proposed by Starobinsky \cite{Staro}, which was based on $R + R^2$ gravity and leads to
a small tilt \cite{MukhChib} and a small value of $r \sim 0.003$. A range of larger values of $r$ are also compatible with the CMB data, 
as are and smaller values, but
the successes of this simple model have led us to explore Starobinsky-like inflationary avatars of no-scale supergravity.
Our first example of this type was based on a minimal Wess-Zumino form of superpotential  \cite{ENO6,Avatars,eno9,ENOV1,ENOV2,king}, but other examples
have emerged  that are based on other forms of superpotential \cite{KLno-scale,Avatars,FKR,FeKR,others,rs,EGNO4,reheating,Moreothers,egnno1,egnno23}. 
In a recent paper we gave a general classification of 
Starobinsky-like inflationary avatars of no-scale supergravity \cite{ENOV1}, and we showed subsequently how such models could be
extended to a unified no-scale model of inflation, supersymmetry breaking and dark energy \cite{ENOV2}.

The no-scale inflationary models described in the previous paragraph were all based on the simplest incarnation of
no-scale supergravity with a K\"ahler potential that could be written in the form \cite{EKN1,EKN2}
\begin{equation}
K \; = \; -3 \, \alpha \, \ln \left(T + T^\dagger - \frac{\phi \phi^\dagger}{3} \right) \, .
\label{3}
\end{equation}
parametrizing a $\frac{SU(2,1)}{SU(2) \cross U(1)}$ coset K\"ahler manifold with the choice $\alpha = 1$,
corresponding to an Einstein space with curvature $R = 2/\alpha = 2$. 
In this case, the prefactor 3 guarantees that in the absence of a superpotential the effective potential vanishes.
Hence it is {\it a fortiori} independent of the field $T$, 
which may correspond to a generic compactification volume modulus, and also $\phi$, which represents a generic chiral matter field.
The no-scale model (\ref{3}) with $\alpha = 1$ was the
starting-point for our recent classification of Starobinsky-like models of inflation~\cite{ENOV1} and their extension to include supersymmetry
breaking and dark energy~\cite{ENOV2}. The possibility of constructing inflationary models with different values of $\alpha$ was first
discussed in~\cite{Avatars}, where it was noted that for $\alpha \lesssim \mathcal{O} (1)$ such models predict
\begin{equation}
n_s \; \simeq \; 1 - \frac{2}{N_*}, \quad r \; \simeq \; \frac{12 \alpha}{N_*^2} \, ,
\label{avatars}
\end{equation}
where $N_*$ is the number of e-folds of inflation. Similar models were later discussed in more detail in~\cite{alpha1,alpha2,att2,rs,att3,att4,otheratt,FK},
where they were termed $\alpha$-attractors. Here we give a general treatment of the construction of no-scale Starobinsky-like
inflationary models with $\alpha \ne 1$.

There are two directions in which the no-scale framework can be generalized: one is to consider K\"ahler manifolds 
that are direct products of irreducible components, i.e., their K\"ahler potentials take the form $K = \sum_n K_n$ where each
$K_n$ is of no-scale type, and the other is to consider no-scale coset manifolds parametrized by multiple chiral fields, e.g., of the forms
$\frac{SU(N,1)}{SU(N) \cross U(1)}$: $N > 2$. In the former case, each of the $K_n$ is of logarithmic form, but they may have different prefactors.
For example, in the case of multiple $\frac{SU(1,1)}{U(1)}$ cosets one may postulate:
\begin{equation}
K \; = \; - \, 3 \, \sum_n \, \alpha_n \, \ln \left(T^n + T_n^\dagger \right) \, ,
\label{several}
\end{equation}
where the quantities $\alpha_n$ are positive, in general. 
Alternatively, the single-coset K\"ahler potential would be generalized to
multiple chiral fields:
\begin{equation}
K \; = \; - \, 3 \, \alpha \, \ln \left(T + T^\dagger - \sum^{N-1}_{i=1} \frac{\phi^i \phi_i^\dagger}{3} \right) \, ,
\label{N}
\end{equation}
and these two options may be combined.

In Witten's original model for string compactification based on the dimensional reduction of 10-dimensional supergravity~\cite{Witten},
there was a single compactification volume modulus $T$ with a K\"ahler potential of the form (\ref{3}) with $\alpha = 1$. However, in a
more general class of string compactifications 
one expects three complex K\"ahler moduli $T_n: n = 1, 2, 3$ with a combined K\"ahler potential of the form (\ref{several})
with $\alpha_1 = \alpha_2 = \alpha_3 = 1/3$. This opens up two additional options for inflation, using either a 
single complex modulus with $\alpha = 1/3$, or a pair linked together as an `area' modulus in a single logarithmic 
K\"ahler potential with an effective prefactor $\alpha = 2/3$. These were the possibilities mentioned in~\cite{Avatars}.

However, there are other chiral fields that one expects to appear in the effective supergravity theory derived from general string theory compactifications.
For example, one generally encounters a complex dilaton/axion field $S$ and some number of complex structure moduli $U^a$.
In general, the dynamical framework for each of these is again a logarithmic non-compact coset K\"ahler potential, and a popular
example is the $STU$ model:
\begin{equation}
K \; = \; - \, \sum_{n=1}^3 \, \ln \left(T^n + T_n^\dagger \right) \,- \;  \ln \left( S + S^\dagger \right) \, - \, \sum_{a=1}^3 \, \ln \left(U^a + U_a^\dagger \right) \, ,
\label{STU}
\end{equation}
in which case the values of $3 \alpha_n$ in (\ref{several}) can take integer values  $\le 7$ 
and can arise if inflation is driven by some linked combination of the $S, T^n$ and $U^a$ fields,
as discussed in~\cite{FK}.

In a previous paper~\cite{ENOV1} we presented a general classification of Starobinsky-like inflationary avatars of $\frac{SU(2,1)}{SU(2) \cross U(1)}$ no-scale supergravity in the case $\alpha = 1$, discussing explicit forms of superpotentials and relations between them. More recently we have shown how supersymmetry breaking and a small cosmological constant could be incorporated in such models~\cite{ENOV2}. In this paper we extend these
constructions to models with $\alpha \ne 1$, and also consider the generalization to models based on $\frac{SU(N,1)}{SU(N) \cross U(1)}$ no-scale supergravity.

The structure of this paper is as follows. In section 2.1, we review briefly the supergravity framework we employ.
We describe the construction of Minkowski and de Sitter 
vacua for models with multiple moduli and matter fields in section 2.2. Some general features of $\alpha$-Starobinsky
models are given in section 2.3. We then go on to consider unified no-scale model, i.e., models that combine Starobinsky-like inflation,
Standard Model (SM)-like phenomenology and
supersymmetry breaking, leaving open the possibility of non-zero and positive vacuum energy after inflation.
We first consider minimal models with a single modulus and a single matter field in section 3.1. While there are many forms
possible for the superpotential giving rise to Starobinsky inflation, they cannot all be implemented in conjunction
with supersymmetry breaking and de Sitter vacua. Indeed, as we show, in many cases when the inflationary
superpotential is perturbed, the minimum of the potential is shifted to a supersymmetry preserving anti-de Sitter (AdS)
vacuum. We generalize these models in section 3.2 by adding multiple moduli and matter fields and allowing for
an arbitrary curvature parameter, $\alpha$. The general construction of  $\alpha$-Starobinsky models
with supersymmetry breaking is developed in section 4. Some discussion of $STU$ models is given in section 5, and
we summarize in section 6.

\section{De Sitter Vacua and No-Scale Attractors}
\subsection{A Brief Review of Supergravity}
We first recall some general features of no-scale supergravity~\cite{LN} that will be useful for our analysis. A generic supergravity theory and its geometric properties are characterized by a K\"ahler potential $K(\Phi^i, {\Phi}^\dagger_j)$, where $\Phi^i$ are the chiral fields, and ${\Phi}^\dagger_j$ are the conjugate fields. Their kinetic energy terms take
the general form:
\begin{equation}
\mathcal{L}_{kin} \; = \; K_i^j \, \partial_{\mu} \Phi^i \partial^{\mu} {\Phi}^\dagger_j,
\label{kin}
\end{equation}
where $K_i^j  \equiv \partial^2 K/ \partial_{\mu} \Phi^i \partial^{\mu} {{\Phi}}^\dagger_j$ is the K\"ahler metric. 
Setting aside the $D$-terms associated with gauge interactions, the $\mathcal{N} = 1$ 
supergravity effective scalar potential is given by (see e.g., \cite{Nilles:1983ge}):
\begin{equation}
V \; = \; e^{G} \left[\pdv{G}{\Phi^i}  {\left(K^{-1} \right)^i_j}  \pdv{G}{{\Phi}^\dagger_j} - 3 \right],
\label{effpot}
\end{equation}
where the K\"ahler function is defined as $G \equiv K + \ln W + \ln W^{\dagger}$, where $K$ is Hermitian and $W$ is holomorphic,
and $\left(K^{-1} \right)^i_j $ is the inverse of the K\"ahler metric. The kinetic energy terms~(\ref{kin}) and the effective scalar potential expression~(\ref{effpot}) 
together yield the corresponding supergravity action for the chiral fields:
\begin{equation}
S \; = \; \int d^4 x \sqrt{-g} \left[ K_i^j \, \partial_{\mu} \Phi^i \partial^{\mu} {\Phi}^\dagger_j - V \right].
\label{act}
\end{equation}
We introduce the following definition of the K\"ahler covariant derivative:
\begin{eqnarray}
\mathcal{D}_i W & \equiv & \partial_i W + K_i W \nonumber \\
\mathcal{D}^i W & \equiv & \partial^i W - K^i W \, ,
\label{cov}
\end{eqnarray}
in terms of which the scalar mass-squared matrix is given by:
\begin{equation}
m_S^2 \; = \; 
\begin{pmatrix}
\left(K^{-1} \right)^i_k \mathcal{D}^k \partial_{j} V & \left(K^{-1} \right)^i_k \mathcal{D}^k \partial^j V \\
\left(K^{-1} \right)^k_i  \mathcal{D}_{k} \partial_{j} V & \left(K^{-1} \right)^k_i  \mathcal{D}_{k} \partial^{{j}} V 
\end{pmatrix},
\label{mass}
\end{equation}
whose diagonalization yields the mass eigenvalues of the scalar fields $\Phi^i$. 
We assume that supersymmetry breaking is generated through an $F$-term that is given by
\begin{equation}
F_i \; = \;  - m_{3/2} \left(K^{-1} \right)_i^j G_j,
\label{fterm}
\end{equation}
where $m_{3/2}$ is the mass of the gravitino, which is given by $m_{3/2} \equiv e^{G/2}$, and  $F$-term supersymmetry breaking is obtained
when $\sum_i |F_i|^2 > 0$ at the minimum.

To understand better the geometric properties of this framework, we recall that
 the K\"ahler curvature for an $\frac{SU(N,1)}{SU(N) \cross U(1)}$ K\"ahler manifold
can be calculated from the following expression~\cite{EKN2}:
\beq
R_i^{j} =  (\log \det G_a^{b})_i^{j} \, ,
\eeq
and the scalar curvature is then
\beq
 \left( {G^{-1}} \right)_j^{i} R_i^{j} \, .
\eeq
For a K\"ahler potential of the form in Eq. (\ref{N}), we find
\beq
R = \frac{N(N+1)}{3 \alpha} \, ,
\eeq
which reduces to the familiar result $R = 2/3$ in minimal $\frac{SU(1,1)}{U(1)}$ no-scale supergravity when $N = 1$ and $\alpha = 1$. 
Note that, because $R$ depends on the total number of chiral fields in the theory,
we cannot determine the K\"ahler curvature $R$ from the cosmological observables $n_s$ and $r$ alone,
though one can in general write
\beq
R \simeq  \frac{N(N+1)(1-n_s)^2}{r } 
\eeq
for $\alpha \lesssim \mathcal{O} (1)$.

\subsection{Minkowski Pair Formulation}
Before considering the generalized construction of
unified no-scale models of inflation, supersymmetry breaking and dark energy, we first show how K\"ahler potentials $K = \sum_n K_n$, where $K_n$ is a K\"ahler potential of no-scale type, can yield de Sitter (dS) vacuum solutions.
We adopt the Minkowski pair formulation that was  first considered in~\cite{enno,ennov}, 
which we use to incorporate an adjustable parameter for supersymmetry breaking and dark energy.

We write the general form of the K\"ahler potential that parametrizes a non-compact $\frac{SU(N,1)}{SU(N) \cross U(1)}$ coset space in the form:
\begin{equation}
K = -3 \, \alpha \ln(\mathcal{V}),
\label{kah1}
\end{equation}
where $\alpha$ is the curvature parameter discussed previously, and we introduce the notation
\begin{equation}
\mathcal{V} \equiv T + T^{\dagger} - \sum_{i=1}^{N-1} \frac{\phi^i \phi_i^{\dagger}}{3}.
\label{arg1}
\end{equation}
As mentioned in the Introduction, the field $T$ may represent a compactification volume modulus and the $\phi_i$ are matter fields. In order to
construct successfully dS vacuum solutions, we assume that the vacuum expectation values (VEVs) of the imaginary field components are fixed to zero, 
i.e., $T^{\dagger} = T$ and $\phi_i^{\dagger} = \phi^i$. This condition can always be achieved dynamically by introducing into
the K\"ahler potential~(\ref{N}) higher-order terms that stabilize the volume modulus field $T$ in the imaginary direction~\cite{EKN3}, as discussed later in this section.

We introduce the following notation:
\begin{equation}
\mathcal{V} \longrightarrow \xi \; \text{when} \; T^{\dagger} = T \; \text{and} \; \phi_i^{\dagger} = \phi^i \, ,
\label{realarg1}
\end{equation}
so that Eq.~(\ref{arg1}) becomes:
\begin{equation}
\xi \; = \;  2T - \sum_{i = 1}^{N - 1} \frac{\phi_i^2}{3}.
\label{realarg2}
\end{equation}
We recover Minkowski vacua solutions by considering the following form of superpotential:
\begin{equation}
W_{M} \; = \; \lambda \cdot \xi^{n_{\pm}},
\label{mink1}
\end{equation}
where $n_{\pm} = \frac{3}{2} \left(\alpha \pm \sqrt{\alpha} \right)$\cite{EKN1}, 
and $\lambda$ is an arbitrary constant. If we combine the pair of Minkowski solutions $\xi^{n_{-}}$ and $\xi^{n_{+}}$
with coefficients $\lambda_{1,2}$, we obtain a superpotential:
\begin{equation}
W_{dS} = \lambda_1 \cdot \xi^{n_{-}} - \lambda_2 \cdot \xi^{n_{+}}
\label{ds1}
\end{equation}
that yields a de Sitter vacuum with the following scalar potential:
\begin{equation}
V = 12 \, \lambda_1 \, \lambda_2 \, .
\label{vac}
\end{equation}
This can be interpreted as the dark energy density, and we recover a positive cosmological constant if $\lambda_1$ and $\lambda_2$ have the same sign. 
In the absence of additional contributions to the vacuum energy density,
the product $\lambda_1 \lambda_2$ should be very small, namely $\mathcal{O} (10^{-120})$ in Planck units. We note that, as 
discussed in~\cite{ENOV2} and seen later,
the difference between $\lambda_1 - \lambda_2$ controls the magnitude of supersymmetry breaking, which is $\gtrsim \mathcal{O} (10^{-16})$ in natural units,
so one might expect a large hierarchy between $\lambda_1$ and $\lambda_2$. However, one does in fact expect additional contributions to the vacuum density
from gauge phase transitions, as we discuss in more detail below.

In this formulation, the potential is given by (\ref{vac}) everywhere in the $N$-dimensional field space corresponding to the real
directions of $T$ and the $\phi_i$.
However, in general there may be instabilities along the imaginary directions. 
The scalar masses in the imaginary direction can be computed using, Eq.~(\ref{mass}) (the masses in the real direction are obviously zero):
\begin{equation}
m_{Im~T}^2 \; = \; \frac{4 \left[\lambda_1^2 (\alpha - 1) \xi^{-3 \sqrt{\alpha}} - 2 (\alpha + 1) \lambda_1 \lambda_2 + \lambda_2^2 (\alpha - 1) \xi^{3 \sqrt{\alpha}} \right]}{\alpha}
\label{mass1}
\end{equation}
and
\begin{equation}
m_{Im~\phi_i}^2  \; = \   \frac{ 4 \left[ \lambda_1^2 (\sqrt{\alpha} - 1) \xi^{-3 \sqrt{\alpha}} + 4 \sqrt{\alpha} \lambda_1 \lambda_2 +  \lambda_2^2 (\sqrt{\alpha} + 1) \xi^{3 \sqrt{\alpha}} \right]}{\sqrt{\alpha}},
\label{mass2}
\end{equation}
where we assume that $\xi \geq 1$. Thus we see that the the imaginary parts of all the matter fields $\phi_i$ acquire identical masses, 
as was to be expected from the symmetric structure of the K\"ahler potential~(\ref{kah1}).

We see in (\ref{mass1}) that when $\alpha < 1$ there is an instability in the imaginary direction of $T$, and (\ref{mass2})
shows that this potentially the case also for the imaginary part of $\phi$. 
These instabilities can be removed by introducing stabilization terms in the K\"ahler potential and the inflationary potential.
For example, one can introduce higher-order stabilization terms in the K\"ahler potential~(\ref{N}) such as postulated in~\cite{EKN3,Avatars}:
\begin{equation}
K \; = \; - \, 3 \, \alpha \, \ln \left[T + T^\dagger + \beta \left(T - T^{\dagger} \right)^4 - \sum^{N-1}_{i=1} \frac{\phi^i \phi_i^\dagger}{3} \right] \quad \text{with}~\beta > 0 \, .
\label{N2}
\end{equation}
The quartic stabilization term $\beta \left(T - T^{\dagger} \right)^4$ does not change the potential in the real direction but stabilizes it in the imaginary direction. 
If, as an example, we fix our fields to $\langle T \rangle = \frac{1}{2}$ and $\langle \phi \rangle = 0$, Eq.~(\ref{mass1}) becomes:
\begin{equation}
m_{Im~T}^2 =\frac{4 \left[\lambda_1^2 (\alpha - 1 + 12 \beta) - 2 (\alpha + 1 - 12 \beta) \lambda_1 \lambda_2 + \lambda_2^2 (\alpha - 1 + 12 \beta) \right]}{\alpha} \, ,
\end{equation}
which can always be stabilized with a suitable choice of $\beta$.

We can extend this framework to more general K\"ahler potential structures, 
in which we consider a combination of $P+1$ no-scale type K\"ahler potentials \cite{ennov},
with a K\"ahler potential of the form:
\begin{equation}
K = - 3 \sum_{n = 1}^{P + 1} \alpha_n \ln (\mathcal{V}_n).
\label{kah2}
\end{equation}
For simplicity, we consider a case in which the first term in this sum contains $N$ matter fields that all belong to a single non-compact 
$\frac{SU(N,1)}{SU(N) \cross U(1)}$ K\"ahler coset manifold, which is denoted as $K_1 = -3 \, \alpha_1 \, \ln(\mathcal{V}_1)$, 
and the remaining $P$ no-scale type K\"ahler potentials are each described by a non-compact $\frac{SU(1,1)}{U(1)}$ coset space. 
Thus, the K\"ahler potential~(\ref{kah2}) becomes:
\begin{equation}
K = - 3 \, \alpha_{1}  \ln (\mathcal{V}_{1}) - 3 \sum_{n = 2}^{P+1} \alpha_n \ln (\mathcal{V}_n),
\label{kah3}
\end{equation}
where:
\begin{equation}
\mathcal{V}_{1} = T + T^{\dagger} - \sum_{i=1}^{N-1} \frac{\phi^i \phi_i^{\dagger}}{3}, \quad \mathcal{V}_n = T^{n} + T_{n}^{\dagger}: \, n> 1,
\label{arg2}
\end{equation}
which parametrizes a non-compact 
$\frac{SU(N,1)}{SU(N) \cross U(1)} \cross \left[\frac{SU(1,1)}{U(1)} \right]^P$ coset space.
We now study its Minkowski vacuum solutions. Just as was done before, 
we restrict our attention to the real directions of the chiral fields, so that,
\begin{equation}
\mathcal{V}_i \longrightarrow \xi_i,~\text{when}~T^{n} = T_{n}^{\dagger} , \quad \phi^i = \phi_{i}^{\dagger}.
\end{equation}
As previously, this requirement can always be achieved dynamically by introducing higher-order stabilization terms, 
as we discuss in more detail when we 
consider multi-field inflationary models in the next section.

As shown in~\cite{ennov}, to recover successfully
Minkowski vacuum solutions for the $\frac{SU(N,1)}{SU(N) \cross U(1)} \cross \left[\frac{SU(1,1)}{U(1)} \right]^P$ 
coset space, we must consider the following superpotential form:
\begin{equation}
W_{M} = \lambda \cdot \prod_{n=1}^{P+1} \xi_n^{\frac{3}{2} \left(\alpha_n - r_{n} \sqrt{\alpha_n} \right)}, \quad \text{with}~ \sum_{n = 1}^{P+1} r_n^2 = 1,
\label{mink2}
\end{equation}
where, as in Eq. (\ref{mink1}), we include only one of the two solutions, $\lambda$ refers to either $\lambda_1$ or $\lambda_2$, 
and $\alpha_n > 0$. Clearly we can interpret the Minkowski superpotential~(\ref{mink2}) as being
specified on the surface of a $P$-sphere with a radius 1. 

In order to obtain de Sitter vacuum solutions for this
$\frac{SU(N,1)}{SU(N) \cross U(1)} \cross \left[\frac{SU(1,1)}{U(1)} \right]^P$ coset space, 
which may be interpreted as solutions with dark energy, we combine two antipodal points lying on the $P$-sphere:
\begin{equation}
W_{dS} = \lambda_1 \cdot \prod_{n=1}^{P+1} \xi_n^{\frac{3}{2}(\alpha_n - r_n \sqrt{\alpha_n})} - \lambda_2 \cdot \prod_{n=1}^{P+1} \xi_n^{\frac{3}{2}(\alpha_n - \bar{r}_n \sqrt{\alpha_n})},
\label{ds2}
\end{equation}
where $r = \left( r_1, r_2, ..., r_{P+1} \right)$, and the antipodal vector is given by $\bar{r} = -\left( r_1, r_2, ..., r_{P+1}  \right) = -r$. 
For convenience, we adopt a notation in which the antipodal vector is written as $\bar{r} = -r$,
so that all our expressions can be expressed in terms of the radial components $r_n$. 
The superpotential~(\ref{ds2}) yields the effective scalar potential $V = 12 \, \lambda_1 \, \lambda_2$ at the extremum
with $\langle T^n \rangle = \frac{1}{2}$ and $\langle \phi^i \rangle = 0$. This extremum might not always be stable,
and its stability should be investigated case by case, but (as discussed previously) the de Sitter minimum can always be stabilized 
in the imaginary direction by introducing the quartic stabilization terms $\beta_n \left(T^n - T_n^{\dagger} \right)^4$ 
in the K\"ahler potential~(\ref{kah2}), a point that we discuss in detail when we combine the dark energy and 
supersymmetry breaking sector with the inflationary sector.

\subsection{$\alpha$-Starobinsky models}

It is well-known that the no-scale formalism we have been discussing is suitable for producing inflationary models
of the Starobinsky type \cite{ENO6,Avatars,eno9,ENOV1,ENOV2,king,KLno-scale,FKR,FeKR,others,rs,EGNO4,reheating,Moreothers,egnno1,egnno23}.
Indeed, for $\alpha = 1$, there are many known examples of superpotentials that lead to a Starobinsky potential for inflation.  
The Wess-Zumino model~\cite{ENO6},
in which the inflaton is associated with a matter field, and the Cecotti model \cite{Cecotti}, in which the inflaton is associated with $T$, are the two 
cases most often considered. It has also been shown~\cite{ENOV1} that these models can be related through the underlying 
$\frac{SU(N,1)}{SU(N) \cross U(1)}$ symmetry of the non-compact no-scale coset space,
and we show later how models with a matter inflaton can be extended to arbitrary values of the
curvature parameter, $\alpha$.

The original Starobinsky model is characterized by the following action:
\begin{equation}
\mathcal{S} \; = \; \frac{1}{2} \int d^4x \sqrt{-g} \left(R + \mu \frac{R^2}{M^2} \right),
\label{act2}
\end{equation}
where $\mu = \frac{1}{6}$ and $M \ll M_P$. One then makes the following Weyl transformation:
\begin{equation}
\tilde{g}_{\mu \nu} = e^{2 \Omega} g_{\mu\nu} =  \left(1 + \frac{\phi}{3M^2} \right) g_{\mu \nu},
\end{equation}
and uses the field redefinition $\phi' = \sqrt{\frac{3}{2}} \ln \left(1 + \frac{\phi}{3 M^2} \right)$.
The action~(\ref{act2}) then becomes:
\begin{equation}
S \; = \; \frac{1}{2} \int d^4 x \sqrt{-\tilde{g}} \left(\tilde{R} -\partial_{\mu} \phi' \partial^{\mu} \phi' - \frac{3}{2} M^2 \left(1 - e^{- \sqrt{\frac{2}{3}} \phi'} \right)^2 
\right),
\label{act3}
\end{equation}
where $\phi'$ is a canonical field. From Eq.~(\ref{act3}) we can see that the Starobinsky inflationary potential will be given by:
\begin{equation}
V \; = \; \frac{3}{4} M^2 \left(1 -e^{- \sqrt{\frac{2}{3 }} \phi'} \right)^2 \,  .
\label{staro}
\end{equation}
It is interesting to note the correspondence between the $R^2$ (de Sitter) and $R + R^2$ (Starobinsky) theories of gravity 
and no-scale supergravity \cite{eno9}. The supergravity Lagrangian written in terms of an Einstein-Hilbert action 
requires a conformal transformation,
which can be expressed in terms of the K\"ahler potential $2\Omega = -K/3$ \cite{cremmer}. 
The details of the correspondence then lie in the choice of the superpotential.
The pure $R^2$ case requires the choice of $W$ given in (\ref{ds1}), 
and the superpotential for the $R + R^2$ models is discussed in the next section.

We note here that the Starobinsky potential can be generalized by changing the K\"ahler curvature to $\alpha \ne 1$. 
In this case, the corresponding conformal transformation is related to the K\"ahler potential by $2\Omega = -K/3\alpha$ and the 
$\alpha$-Starobinsky scalar potential
becomes (again with a suitable choice of superpotential) 
\begin{equation}
V \; = \; \frac{3}{4} M^2 \left(1 -e^{- \sqrt{\frac{2}{3 \alpha}} \phi'} \right)^2 \, .
\label{astaro}
\end{equation}
The cosmological observables for $\alpha$-Starobinsky potential~(\ref{astaro}) are given by~(\ref{avatars}). 
Our goal in the following is to unify no-scale models that incorporate the $\alpha$-Starobinsky inflationary model
(or a general no-scale attractor inflationary potential) at a scale $\mathcal{O}(10^{13})$ GeV, 
with an adjustable scale for supersymmetry breaking 
and a cosmological constant $\mathcal{O}(10^{-120})$.
To achieve such unification, we must stabilize strongly the volume modulus $T$, 
an aspect that is deferred to the next section. 

\begin{figure}[ht!]
\centering
\includegraphics[scale=0.6]{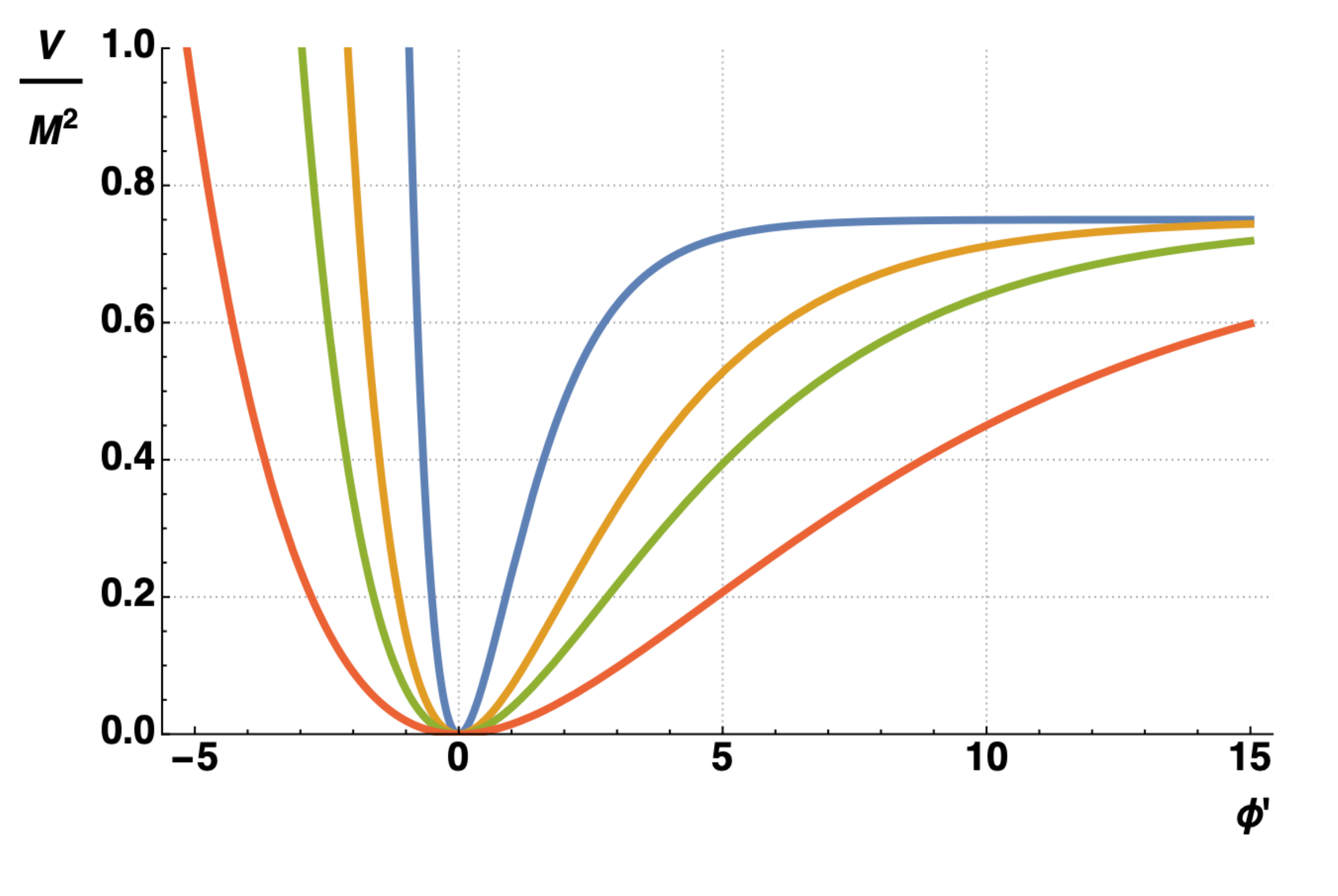}
\caption{\it The $\alpha$-Starobinsky potentials for different values of the curvature parameter $\alpha$. 
The blue line corresponds to the original Starobinsky inflationary potential with $\alpha = 1$, 
the yellow line corresponds to $\alpha = 5$, the green line corresponds to $\alpha = 10$, 
and the orange line corresponds to $\alpha = 30$.}
\label{staroplot}
\end{figure}

For illustration, we plot $\alpha$-Starobinsky potential forms with different values of $\alpha$ in Fig.~\ref{staroplot}. 
We can see from the Figure that increasing the value of the curvature parameter 
$\alpha$ stretches the Starobinsky potential horizontally, reducing the flatness of the plateau at any fixed value of $\phi^\prime$.

\begin{figure}[ht!]
\centering
\includegraphics[scale=0.55]{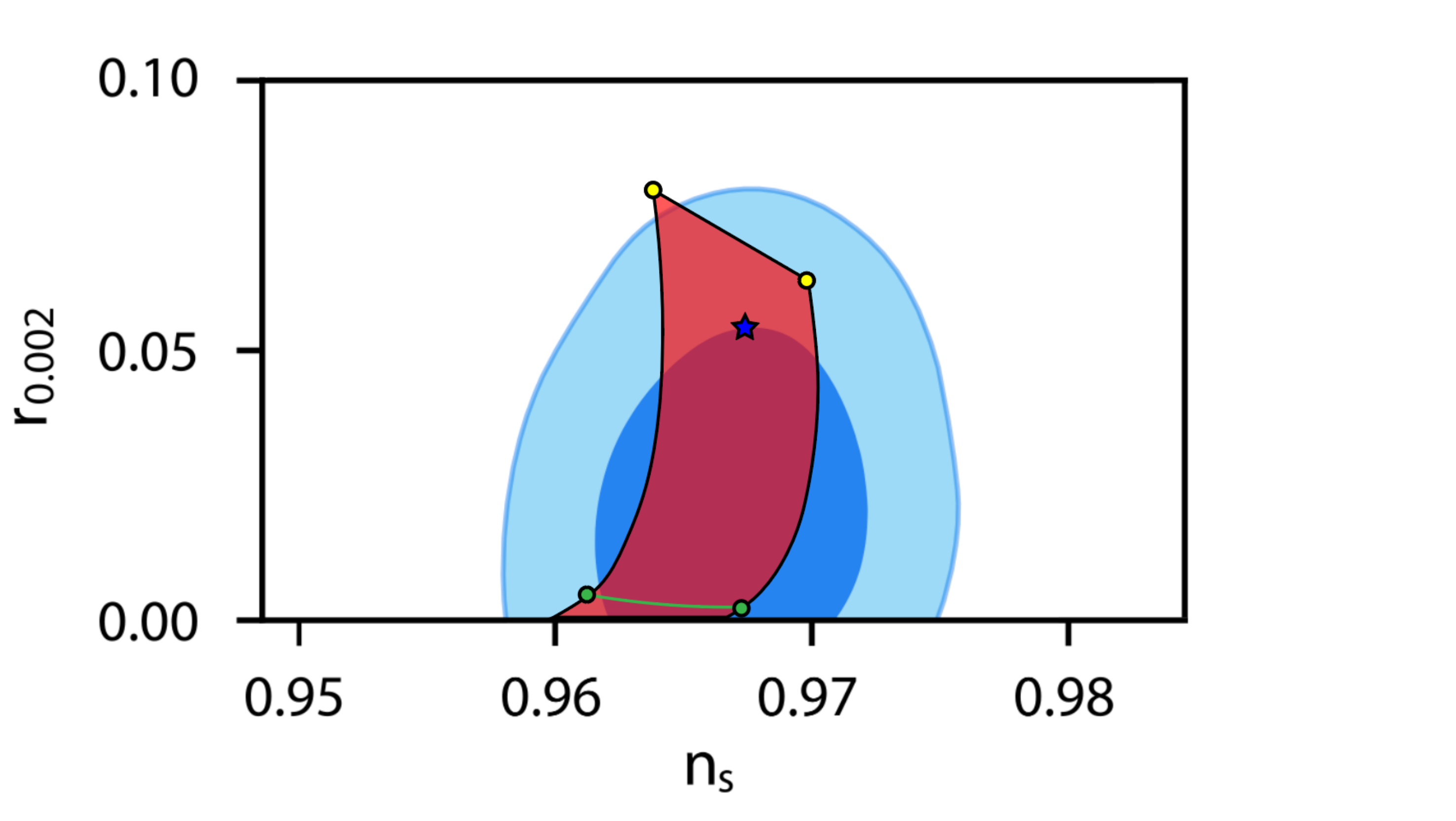}
\caption{\it The cosmological observables $n_s$ and $r$ for the $\alpha$-Starobinsky potential for $N_*$ between 50 (left) and 60 (right). 
The upper (lower) pair of yellow (green) dots are the predictions when $\alpha = 100$ ($\alpha  = 1$, corresponding to the Starobinsky model), while the lower end of the swath represents the 
cosmological observables when $\alpha \rightarrow 0$.  
The blue shadings correspond to the 68\% and 95\%
confidence level regions from Planck data combined with BICEP2/Keck and BAO results~\cite{planck18} The 68\% upper bound $r \lesssim 0.055$, indicated by the blue star, is attained
for $\alpha \sim  51$ when $n_s \sim 0.967$, for a nominal choice of $N_* \approx 55$.}
\label{staroplot2}
\end{figure}

We show in Fig.~\ref{staroplot2} the predicted cosmological observables for these $\alpha$-Starobinsky potentials in the $(n_s, r)$ plane,
together with the results of the Planck collaboration combined with other CMB data,
indicated by blue shadings corresponding to the 68\% and 95\%
confidence level regions~\cite{planck18}~\footnote{Note that the $\alpha$-Starobinsky predictions
include corrections beyond the small-$\alpha$
values shown in (\ref{avatars}), since they were calculated by numerical integration of the equation of motion for the inflaton field. Similar predictions were shown in Fig.~1 of~\cite{alpha1}, but corrections of higher order in $\alpha$
are absent from the corresponding $\alpha$-attractor predictions in Fig.~1 of~\cite{FK} and Fig.~8 of~\cite{planck18}.}.
As the curvature parameter $\alpha$ increases, the value of the scalar tilt $n_s$ changes only 
slightly and stays within the range $\sim 0.96-0.97$, while the tensor-to-scalar ratio $r$ increases with the value of $\alpha$.
The CMB data set a a 68\% upper bound on the tensor-to-scalar ratio $r \sim 0.055$, which is attained
for $\alpha \sim  51$ when $n_s \sim 0.967$ for a nominal choice of $N_* \approx 55$,
as indicated by the blue star. The green dots and line at small $r$ show the prediction of the original Starobinsky model, corresponding to the case $\alpha = 1$.
It is apparent that future measurements of $r$ will be able to constrain $\alpha$ more significantly, 
and that more precise measurements of $n_s$
could in principle constrain $n_s$, thereby $N_*$ and hence the post-inflationary history of the Universe, which is
sensitive to the decay of the inflaton into low-mass particles~\cite{reheating}.

\section{Unified No-Scale Models}
\subsection{Supersymmetry Breaking and Inflationary Dynamics}
We now show how one can combine the dark energy sector $W_{dS}$ with various inflationary models to obtain a 
unified no-scale supergravity model. Because we want to consider 
Minkowski pair models that break supersymmetry, 
we impose the condition that $W_{I}$ does not break the supersymmetry at the minimum. 
For simplicity, we consider models based on a non-compact $\frac{SU(2,1)}{SU(2) \cross U(1)}$ coset space 
with curvature parameter set to $\alpha = 1$, i.e., the K\"ahler potential given in Eq. (\ref{3}).
In order to incorporate supersymmetry breaking via the Minkowski pair superpotential $W_{dS}$, 
we need to consider specific inflationary superpotential forms 
that do not lead to unphysical supersymmetry-preserving AdS vacua states. 
It can readily be shown that the Minkowski pair construction combined with an arbitrary superpotential
that is a function of the volume modulus $T$ only, 
$W_I = f(T)$, can give dS vacuum states with broken supersymmetry. 
However, we do not consider such models here because, as discussed in~\cite{Avatars},
superpotentials of this form cannot lead to Starobinsky-like inflationary potentials with unbroken supersymmetry at the minimum.

It was shown in~\cite{ENO6} that the Starobinsky model of inflation can be obtained from the Wess-Zumino form of superpotential, 
which is a function of the matter-like field $\phi$ only. Thus, we are led to consider the following inflationary superpotential form:
\begin{equation}
W_{I} = f(\phi),
\label{supphi}
\end{equation}
where we assume that the minimum is located at $\langle \phi \rangle = 0$. 
 The superpotential~(\ref{supphi}) yields the following compact form of scalar potential:
\begin{equation}
V = \frac{f'(\phi)^2}{\left(2T - \frac{\phi^2}{3} \right)^2}\, .
\end{equation}
and requiring vanishing vacuum energy at the minimum imposes the condition $f'(0) = 0$.
If we also require the minimum to be supersymmetric, we have $D_T W_I = \partial_T{W_I} + K_T W_I  \simeq  - 3 f(\phi)  = 0$,
which implies that  $f(0) = 0$~\footnote{We note that $D_\phi W = 0$ is automatically satisfied if $\phi = 0$ and $f'(0) = 0$ at the minimum.}.

We can now combine the inflaton superpotential~(\ref{supphi}) with the Minkowski pair superpotential $W_{dS}$, obtaining
\begin{equation}
W \; = \; W_{I} + W_{dS} = f(\phi) + \lambda_1 - \lambda_2 \left(2T - \frac{\phi^2}{3} \right)^3,
\end{equation}
which yields the following effective scalar potential:
\begin{equation}
V \; = \; 12 \lambda_1 \lambda_2 + 12 \lambda_2 f(\phi)  + \frac{f'(\phi)^2}{\left(2T - \frac{\phi^2}{3} \right)^2}.
\label{phipot}
\end{equation}
Taking derivatives with respect to the fields $T$ and $\phi$, we obtain
\begin{equation}
V_T \; = \; -\frac{4 f'(\phi )^2}{\left(2 T-\frac{\phi ^2}{3}\right)^3}
\end{equation}
and 
\begin{equation}
V_{\phi} \; = \; 6 f'(\phi ) \left(\frac{3 \left(2 T-\frac{\phi ^2}{3}\right) f''(\phi )+2 \phi  f'(\phi )}{9 \left(2 T-\frac{\phi ^2}{3}\right)^3}+2 \lambda _2\right).
\end{equation}
We can see from these expressions that, as long as the condition $f'(0) = 0$ is satisfied,
the position of the minimum does not shift when the dark energy superpotential $W_{dS}$ is introduced. 
In addition, at the minimum the effective scalar potential~(\ref{phipot}) reduces to the dS vacua solutions $V = 12 \, \lambda_1 \, \lambda_2$.
While the $F$-term for $\phi$ remains zero (when $\phi = 0$ and $f'(0) = 0$ at the minimum), the $F$-term for $T$ is proportional to 
$D_T W = -3 \, (f(\phi) + \lambda_1 + \lambda_2) = -3(\lambda_1 + \lambda_2)$, indicating that supersymmetry is
broken. Thus we have shown that supersymmetry breaking with a positive cosmological constant 
can be achieved if one considers an inflationary superpotential of the form $W_{I} = f(\phi)$.

However, the dark energy sector cannot be combined with a general superpotential of the form $W_{I} = F(T, \phi)$. 
To illustrate that, we consider the following superpotential form:
\begin{equation}
W_{I} = F(T, \phi) = f(T) \cdot \phi.
\label{suptandphi}
\end{equation}
We obtain from~(\ref{suptandphi}) the following scalar potential:
\begin{equation}
V = \frac{\frac{2}{3} T \phi ^2 f'(T)^2-\frac{4}{3} \phi ^2 f(T) f'(T)+f(T)^2}{\left(2 T-\frac{\phi ^2}{3}\right)^2},
\end{equation}
whose derivative with respect to $\phi$ is given by:
\begin{equation}
V_{\phi} = \frac{4 \phi  \left(-6 f(T) \left(2 T+\frac{\phi ^2}{3}\right) f'(T)+3 T \left(2 T+\frac{\phi ^2}{3}\right) f'(T)^2+3 f(T)^2\right)}{9 \left(2 T-\frac{\phi ^2}{3}\right)^3}.
\end{equation}
We see from this that the minimum must be located at $\langle \phi \rangle = 0$, and we assume that the minimum is also at 
$\langle T \rangle = 1/2$. Using this condition, we obtain the following expression for the derivative with respect to $T$ at the minimum:
\begin{equation}
V_T \; = \; 2 f \left(1/2 \right) \left[ f' \left(1/2 \right) - 2 f \left( 1/2 \right) \right],
\end{equation}
which shows that at the minimum we must also satisfy the condition $f(1/2) = 0$, which also guarantees $V=0$ at the minimum.

Next, we combine the inflationary superpotential~(\ref{suptandphi}) with the dark energy sector:
\begin{equation}
W = W_{I} + W_{dS} = f(T) \cdot \phi + \lambda_1 - \lambda_2 \left(2T - \frac{\phi^2}{3} \right)^3,
\end{equation}
which yields the following effective scalar potential:
\begin{eqnarray}
V & = & \frac{1}{\left(2 T-\frac{\phi ^2}{3}\right)^2} \left[ \frac{2}{3} T \phi ^2 f'(T)^2+12 \lambda _2 \left(2 T-\frac{\phi ^2}{3}\right)^2 \left(\phi  f(T)+\lambda _1\right)
\right. \nonumber \\
& & \left.
+f(T)^2 -2 \phi  f'(T) \left(\frac{1}{3} \left(2 \phi  f(T)+3 \lambda _1\right)+\lambda _2 \left(2 T-\frac{\phi ^2}{3}\right)^3\right) \right].
\label{potfull}
\end{eqnarray}
The minimum of this more complicated scalar potential is 
shifted to a new position. To find this shift, we consider small perturbations around the initial minimum, given by:
\begin{equation}
f^{(n)} \left( 1/2 + \delta t \right) \approx f^{(n)} \left(1/2 \right) + f^{(n+1)}(1/2) \cdot \delta t , \quad~\text{with}~n =0, 1, 2,
\end{equation}
and
\begin{equation}
\langle \phi \rangle + \delta \phi = \delta \phi.
\end{equation}
Using these perturbations in the scalar potential~(\ref{potfull}), we find:
\begin{equation}
\delta \phi \; \approx \; \frac{3 \left( \lambda_1 + \lambda_2 \right)}{f'(1/2)}
\qquad
\delta T \; \approx \; - \frac{3\left( \lambda_1 + \lambda_2 \right)^2}{f'(1/2)^{2}}, 
\label{approx1}
\end{equation}
which yields the following effective scalar potential at the minimum:
\begin{equation}
V \approx -3 \left(\lambda_1 - \lambda_2 \right)^2,
\label{adspot}
\end{equation}
We see from~(\ref{adspot}) that, when $\lambda_1 \neq \lambda_2$,
 the minimum always shifts to a supersymmetry-preserving AdS vacuum, at least for small $\lambda_{1,2}$.

The Cecotti form of the superpotential $W_I = \sqrt{3} M \phi (T-1/2)$ \cite{Cecotti} falls into the
category of superpotentials that is not suitable for combination with
our Minkowski pair formulation of broken supersymmetric dS solutions. 
We argued previously~\cite{ENOV1} for an equivalence among the many avatars of superpotentials yielding 
Starobinsky inflation based on an the underlying SU(2,1) invariance.  However, when $W_{dS}$ is added to the
theory, this invariance is broken and the avatars are no longer equivalent.
Therefore, in the remainder of the paper, we combine the dark energy superpotential $W_{dS}$ with inflationary superpotentials 
that are functions of matter fields only. More general forms for $W(T,\phi)$ may be possible, but we have not explored the
general conditions on $W$ and its derivatives for models other than the three sets of models that have attracted most interest in this context. 

We note that the inflationary superpotential $W_{I}$ contains a single parameter, denoted by $M$,
which is {\em not} of order unity. In the Wess-Zumino model discussed in more detail below, $M$ corresponds to the
inflaton mass and its magnitude is set by the normalization of the scalar density perturbations~\cite{planck18}, so that 
$ M \simeq 1.2 \times 10^{-5} M_P \simeq 3 \times 10^{13}$ GeV. In contrast, the constants in $W_{dS}$ must be significantly smaller.
Supersymmetry breaking is characterized by an $F$-term of the form in  Eq. (\ref{fterm}) and is given by:
\begin{equation}
\sum_{i = 1}^2  |F_i|^2 = F_T^2 = \frac{\left(\lambda_1 + \lambda_2\right)^2 }{\alpha}\, ,
\end{equation}
where the supersymmetry breaking is generated through an $F$-term for $T$. For $\alpha = 1$, $F_T = (\lambda_1 + \lambda_2)$, and
the gravitino mass is given by:
\beq
m_{3/2} = e^{G/2} = e^{K/2} \; W = (\lambda_1 - \lambda_2)  
\eeq
at the minimum, and is independent of $\alpha$. Thus we expect the difference of the two parameters to be of order $10^{-16}$ in Planck units. 
Hence the terms in potential~(\ref{phipot}) that are coupled to $\lambda_{1,2}$ can be safely neglected during inflation, 
and do not affect the slow-roll dynamics.

We recall that the vacuum energy density at the minimum is given by  $\Lambda = 12 \lambda_1 \lambda_2$, 
Thus, in the absence of any phase transitions, we must require $12 \lambda_1 \lambda_2 \sim \mathcal{O} (10^{-120})$,
which is possible if one of the two constants is hierarchically much smaller than the other, e.g., $\lambda_1 \sim 10^{-15}$ and 
$\lambda_2 \sim 10^{-105}$. 
However, we know that the vacuum energy today (i.e., the cosmological constant) is a sum of contributions
that have changed during phase transitions
throughout the history of the Universe. 
For example, the electroweak transition would make a contribution $\sim - {\cal O}(10^{-60})$, which could be cancelled
to a sufficiently small value if $\lambda_1 \sim 10^{-15}$ and 
$\lambda_2 \sim 10^{-45}$. However, a grand unified (GUT) transition would plausibly make a contribution $\sim - {\cal O}(10^{-30})$,
corresponding to vacuum energy of 
$\sim 10^{11} $ GeV as is typical in flipped SU(5) models where the vacuum energy is related to $m_{3/2}^2 M_{GUT}^2$ (see, e.g., \cite{EGNNO3})
which could be cancelled by $\lambda_1 \sim \lambda_2 \sim 10^{-15}$ to provide a suitable cosmological constant today,
Without loss of generality, we can define:
\beq
\lambda_1 \, = \, {\tilde \lambda}_1 M^3 \qquad \lambda_2\ \, = \, {\tilde \lambda}_2 M^3 \, ,
\label{ltilde}
\eeq
and we return below to the possibility of a cancellation when we discuss the Wess-Zumino model in more detail. 

\subsection{Multi-field No-Scale Attractors with Supersymmetry Breaking}
It is relatively straightforward to generalize the previous results to multi-field no-scale attractors.
We consider the K\"ahler potential~(\ref{kah3}) that parametrizes a non-compact 
$\frac{SU(N,1)}{SU(N) \cross U(1)} \cross \left[\frac{SU(1,1)}{U(1)} \right]^P$ coset space, 
and combine it with the dS vacuum solutions~(\ref{ds2}). The matter fields $\phi^i$, 
which will be associated with inflaton fields, are described by the no-scale 
K\"ahler potential $K_1 = -3 \alpha_1 \ln(\mathcal{V}_1)$, where the function $\mathcal{V}_1$ is given by~(\ref{arg1}). 
The multi-field no-scale attractors will be characterized by a curvature parameter $\alpha_1$. 

As discussed in the previous Section, we assume that the superpotential associated with inflation is a  function of matter fields only. 
Thus, the unified superpotential can be expressed as:
\begin{equation}
W = W_{I} + W_{dS} = f(\bm{\phi}) + \lambda_1 \cdot \prod_{n=1}^{P + 1} \xi_n^{\frac{3}{2}(\alpha_n - r_n \sqrt{\alpha_n})} - \lambda_2 \cdot \prod_{n=1}^{P + 1} \xi_n^{\frac{3}{2}(\alpha_n + r_n \sqrt{\alpha_n})},
\label{unisup1}
\end{equation}
where $W_I = f(\bm{\phi})$, and $\bm{\phi} \equiv \{ \phi_1, \phi_2, ..., \phi_{N-1} \}$. 
Exactly as before, we require that at the minimum $W_{I} = f(0) = 0$ and $f'(0) = 0$. 
We assume that the chiral fields at the minimum obtain vacuum expectation values $\langle T^n \rangle = \frac{1}{2}$ 
and $\langle \phi^i \rangle = 0$, and in this case the gravitino mass becomes $m_{3/2} = \lambda_1 - \lambda_2$,
also as before, while the $F$-term giving rise to supersymmetry breaking~(\ref{fterm}) is given by:
\begin{equation}
\sum_{i=1}^{P + 1} |F_i|^2 =  \sum_{i=1}^{P + 1} \frac{r_i^2}{ \alpha_i} \left(\lambda_1 + \lambda_2 \right)^2 \, .
\label{fbreak}
\end{equation}
Note that we have only included the sum over moduli in (\ref{fbreak}). It is relatively easy to see that the $F$-terms
associated with matter fields (including the inflaton) are all zero.  
Next, we use the K\"ahler potential expression~(\ref{kah3}) with the superpotential form~(\ref{unisup1}), and 
obtain the following effective scalar potential:
\begin{equation}
\begin{split}
& V =  12 \lambda_1 \lambda_2 + 3 f(\bm{\phi})^2 \left(\sum_{n=1}^{P+1} \alpha_n - 1 \right)\prod_{n = 1}^{P + 1} \xi_n^{-3 \alpha_n} + \frac{\prod_{n=1}^{P + 1} \xi_n^{-3 \alpha_n}}{\alpha_1} \cdot \xi_1 \cdot  \sum_{i = 1}^{N - 1} \partial_i f(\bm{\phi})^2 \\
& - 6 f(\bm{\phi}) \cdot \left[ \lambda_1 \left(1 + \sum_{n=1}^{P+1} r_n \sqrt{\alpha}_n \right) \cdot \prod_{n =1}^{P + 1} \xi_n^{- \frac{3}{2} \left(\alpha_n - r_n \sqrt{\alpha_n} \right)} -  \lambda_2 \left(1 - \sum_{n=1}^{P + 1} r_n \sqrt{\alpha}_n \right) \cdot \prod_{n =1}^{P+1} \xi_n^{- \frac{3}{2} \left(\alpha_n + r_n \sqrt{\alpha_n} \right)}\right],
\end{split} 
\label{unipot1}
\end{equation}
where $\partial_i f(\bm{\phi}) \equiv \frac{\partial f(\bm{\phi})}{\partial \phi_i}$.
We can identify four distinct contributions to to $V$. The first term is once again the vacuum energy density after inflation. 
The second term is proportional to $f^2$ and is potentially dangerous, as it could seriously impact the inflaton potential.
The third term is the generalization of the inflaton potential, and the final term is related to the supersymmetry-breaking terms.
These terms can be neglected during inflation, as $\lambda_i \ll M$. 

In order to safeguard Starobinsky-like inflation, we must ensure the absence of the second term. 
There are two ways to achieve this. The first is rather obvious, namely we could require 
\beq
\sum_{n=1}^{P+1} \alpha_n = 1 \, .
\eeq
However, there is a more elegant (and general) solution to this problem, which at the same time simplifies the scalar potential. 
The key is to couple our inflationary potential to the Minkowski vacuum solution, given by~(\ref{mink2}). Thus, we consider the following superpotential form:
\begin{equation}
W = W_{dS} + W_{M} \cdot W_{I},
\end{equation}
which amounts to adding $f(\bm{\phi})$ to either $\lambda_1$ or $\lambda_2$ in Eq. (\ref{mink2}). 
Thus, we can couple the inflationary superpotential in two different ways: either
\begin{equation}
W =( f(\bm{\phi}) + \lambda_1 )\cdot \prod_{n = 1}^{P + 1} \xi_n^{\frac{3}{2} (\alpha_n - r_n \sqrt{\alpha_n})} - \lambda_2 \cdot \prod_{n=1}^{P + 1} \xi_n^{\frac{3}{2}(\alpha_n + r_n \sqrt{\alpha_n})},
\label{form1}
\end{equation}
which yields the following scalar potential:
\begin{equation}
V = 12 \lambda_1 \lambda_2 + 12 \lambda_2 f(\bm{\phi})  + \frac{\prod_{n =1}^{P+1} \xi_n^{-3 \alpha_n}}{\alpha_1} \cdot \xi_1 \cdot \sum_{i = 1}^{N - 1} \partial_i f(\bm{\phi})^2,
\label{unipot2}
\end{equation}
or as a second possibility:
\begin{equation}
W =\lambda_1 \cdot \prod_{n=1}^{P+1} \xi_n^{\frac{3}{2} (\alpha_n - r_n \sqrt{\alpha_n})} -(f(\bm{\phi}) + \lambda_2) \cdot \prod_{n=1}^{P+1} \xi_n^{\frac{3}{2}(\alpha_n + r_n \sqrt{\alpha_n})},
\end{equation}
which gives the following scalar potential:
\begin{equation}
V = 12 \lambda_1 \lambda_2 + 12 \lambda_1 f(\bm{\phi}) + \frac{\prod_{n =1}^{P+1} \xi_n^{-3 \alpha_n}}{\alpha_1} \cdot \xi_1 \cdot \sum_{i = 1}^{N - 1} \partial_i f(\bm{\phi})^2.
\label{unipot3}
\end{equation}
We note that the scalar potentials (\ref{unipot2}) and (\ref{unipot3}) have relatively simple forms.
Once again, the first term in each equation is the vacuum energy density after inflation,
and the second term, while proportional to $f(\bm{\phi})$, is rendered
harmless as it is proportional to one of the two small coefficients $\lambda_i$. The third term in each case
leads to $\alpha$-Starobinsky inflation. 
We need only impose the conditions that $f(\bm{\phi})$ and $f'(\bm{\phi})$ are zero at the minimum.

As mentioned previously, to stabilize the moduli fields $T^n$ dynamically at their 
vacuum expectation values to $\langle T^n \rangle = \frac{1}{2} $, 
we introduce quartic stabilization terms \cite{EKN3, Avatars} in the K\"ahler potential. Thus, 
the complete K\"ahler potential based on non-compact $\frac{SU(N,1)}{SU(N) \cross U(1)} \cross \left[ \frac{SU(1,1)}{U(1)} \right]^P$ 
coset space takes the form:
\begin{eqnarray}
K & = & -3 \alpha_1   \ln  \left[T + T^{\dagger} + \beta^R \left(T + T^{\dagger} - 1 \right)^4 + \beta^I \left(T - T^{\dagger} \right)^4 - \sum_{i=1}^{N-1} \frac{|\phi^i|^2}{3} \right]  \nonumber \\
&& - 3 \sum_{n = 2}^{P+1} \alpha_n   \ln \left[T^{n} + T_n^{\dagger} + \beta_n^R \left(T^{n} + T_n^{\dagger}-1\right)^4 + \beta_n^I \left(T^{n} - T_n^{\dagger}\right)^4 \right].
\label{kah4}
\end{eqnarray}
Inflation is described by the generalization of $-3 \alpha_1 \ln(\mathcal{V}_1)$, where we assume that inflation is driven by the matter fields $\phi^i$. 
The term proportional to $\beta^R$ fixes $\langle T \rangle = \frac{1}{2}$ as needed to generate the Starobinsky potential. 
Since the potential is actually flat along the real $T^n$ directions, we use 
the terms proportional to the $\beta_n^R$ to fix the remaining real parts of the moduli to the same value~\footnote{The actual fixed values
are unimportant, and can be fixed to a set of constants $c_n$.}.
Hence, during inflation, we obtain $\xi_n = 1$ for $n \geq 2$, 
and the scalar potential forms~(\ref{unipot2}) and~(\ref{unipot3}) can be approximated by:
\begin{equation}
V \approx \frac{\sum_{i = 1}^{N - 1} \partial_i f(\bm{\phi})^2}{\alpha_1 \, \xi_1^{3 \alpha_1 - 1}},
\label{unipot4}
\end{equation}
which, after fixing the volume modulus to $\langle T \rangle = \frac{1}{2}$, 
becomes:
\begin{equation}
V \approx \frac{\sum_{i = 1}^{N - 1} \partial_i f(\bm{\phi})^2}{\alpha_1 \left(1 - \sum_{i=1}^{N - 1} \frac{\phi_i^2}{3} \right)^{3 \alpha_1 - 1}}.
\label{unipot5}
\end{equation}
The next step is to obtain the kinetic terms for our unified multi-field no-scale attractor models.
After setting the moduli to their vacuum values, the K\"ahler potential~(\ref{kah3}) yields
\begin{equation}
\mathcal{L}_{kin} = \frac{\alpha_1}{\left(1 - \sum_{i =1}^{N-1} \frac{\phi_i^2}{3} \right)^2} \cdot \left[ \sum^{N+1}_{i = 1} \left(1 - \sum_{j = 1, j \neq i}^{N -1}\frac{\phi_j^2}{3} \right) \cdot (\partial_{\mu} \phi_i)^2 
+ \frac{2}{3} \sum_{i, j = 1; i \neq j}^{N-1} (\phi_i \, \partial_{\mu} \phi_i)
\cdot (\phi_j \, \partial_{\mu} \phi_j)
\right].
\label{kinatt}
\end{equation}
Although the kinetic terms of the Lagrangian~(\ref{kinatt}) may appear complicated, the Lagrangian is still highly symmetric.

\section{No-Scale $\alpha$-Starobinsky Models with Supersymmetry Breaking}

In this Section, we examine in more detail no-scale $\alpha$-Starobinsky models
characterized by a non-compact $\frac{SU(2, 1)}{SU(2) \cross U(1)}$ K\"ahler potential~(\ref{3}).
 As discussed in the previous Section, we assume that the no-scale K\"ahler potential is modified to include a quartic 
stabilization term which fixes the VEV of the volume modulus $\langle T \rangle = \frac{1}{2}$. During inflation,
 the imaginary part of $\phi$ picks up a mass, so
we can take $\langle Im~\phi \rangle = 0$, and we associate the real part of the field $\phi$ with the inflaton. 
We make the following field redefinition to obtain a canonically-normalized field:
\begin{equation}
\phi = \sqrt{3} \tanh(\frac{x}{\sqrt{6 \alpha}}).
\label{kinphi}
\end{equation}
It was shown in~\cite{ENO6} for $\alpha = 1$ that  Starobinsky inflation can be derived from an 
$\frac{SU(2,1)}{SU(2) \cross U(1)}$ K\"ahler potential with a Wess-Zumino superpotential:
\begin{equation}
W_{I} = M \left(\frac{\phi^2}{2} - \frac{\phi^3}{3 \sqrt{3}} \right) \, ,
\label{wi}
\end{equation}
which leads to the effective scalar potential given in Eq. (\ref{staro}) with the replacement $\phi' \to x$. 

The Wess-Zumino superpotential (\ref{wi}) can be combined with the supersymmetry breaking and dark energy sector $W_{dS}$ using Eq. (\ref{ltilde}):
\begin{equation}
W = W_{I} + W_{dS} = M \left(\frac{\phi^2}{2} - \frac{\phi^3}{3 \sqrt{3}} \right) + {\tilde \lambda}_1 M^3 - {\tilde \lambda}_2 M^3 \left(2T - \frac{\phi^2}{3} \right)^3,~\text{for}~\alpha = 1.
\label{uniwz}
\end{equation}
The unified Wess-Zumino model~(\ref{uniwz}) with the fields fixed at $\langle T \rangle = \frac{1}{2}$ and $\langle \Im~\phi \rangle = 0$
then yields the following scalar potential:
\begin{equation}
V \; = \; 12  {\tilde \lambda}_1 {\tilde \lambda}_2 M^6 + 12 {\tilde \lambda}_2  M^4 \left(\frac{\phi^2}{2} - \frac{\phi^3}{3 \sqrt{3}} \right) + 3M^2 \left(\frac{\phi}{\sqrt{3} + \phi} \right)^2,
\label{potphi}
\end{equation}
which, after canonical field redefinition~(\ref{kinphi}), becomes:
\begin{equation}
V \; = \; 12  {\tilde \lambda}_1 {\tilde \lambda}_2 M^6 + 6 {\tilde \lambda}_2 M^4 \tanh^2 \left(\frac{x}{\sqrt{6}} \right) \left(3 - 2 \tanh(\frac{x}	{\sqrt{6}}) \right) + \frac{3}{4}M^2 \left(1 - e^{- \sqrt{\frac{2}{3}} x} \right)^2.
\label{unistaro1}
\end{equation}
The first term in (\ref{unistaro1}) corresponds to the cosmological constant, $\Lambda = 12 {\tilde \lambda}_1 {\tilde \lambda}_2  M^6$.  
As mentioned earlier, we expect that the vacuum energy density is modified by (negative) contributions from
phases transitions occurring after inflation. For example, for ${\tilde \lambda}_{1,2} \sim \mathcal{O}(1)$,
we would require a contribution of order $M^6 \sim 10^{-30}$ to cancel the term in (\ref{unistaro1})
to eventually yield a cosmological constant of order $10^{-120}$ today. Interestingly,
the GUT phase transition in a flipped SU(5) $\times$ U(1) model occurs after inflation \cite{EGNNO3} and contributes
$\Delta V \sim - M_{\rm susy}^2 M_{\rm GUT}^2 \sim - ({\tilde \lambda}_1 - {\tilde \lambda}_2)^2 M^6 M_{\rm GUT}^2$ and would indicate that perhaps ${\tilde \lambda}_1/{\tilde \lambda_2} \sim  (M_{\rm GUT}/M_P)^2$ or equivalently for 
${\tilde \lambda}_2/{\tilde \lambda_1}$.

The second term in (\ref{unistaro1}) corresponds to a perturbation of the inflaton potential and
 has a negligible effect on the inflationary dynamics, because it is scaled by $M^4$ relative to the 
 inflationary potential (the third term in (\ref{unistaro1})) which scales as $M^2$. 
 Therefore, at large $x$ $\Delta V$ adds a relatively small amount $6 {\tilde \lambda}_2 M^4$ to the Starobinsky
plateau value of $(3/4) M^2$.

Our next goal is to extend this formalism to general cases with $\alpha \neq 1$,
and construct generalized $\alpha$-Starobinsky inflationary models consistent with supersymmetry breaking 
and a positive cosmological constant. We follow the treatment presented in the previous Section and, for simplicity, 
consider only the cases where the inflationary superpotential $W_{I}$ is coupled to a 
Minkowski vacuum solution associated with the lower power $\xi^{n_{-}}$. Our goal therefore, is to determine
the superpotential which generates a Starobinsky potential for any value of $\alpha$.  To this end,
we parametrize the superpotential with a function $f(\phi)$ as:
\begin{equation}
W_{I} \; = \; \sqrt{\alpha} \, f(\phi) \cdot \left(2T - \frac{\phi^2}{3} \right)^{\frac{3}{2} \left(\alpha - \sqrt{\alpha} \right)}. 
\label{Wf}
\end{equation}
In the real direction ($ \phi = \phi^\dagger$ and $T = T^\dagger$), this reduces to the relatively simple form:
\begin{equation}
V = \left(2 T-\frac{\phi ^2}{3}\right)^{1-3 \sqrt{\alpha }} \cdot f'(\phi )^2 \, ,
\label{genstaro}
\end{equation}
where $f'(\phi) = df/d\phi$.
Setting the potential (\ref{genstaro}) to the Starobinsky potential, e.g., as in the third term of Eq. (\ref{potphi}),
we can determine $f(\phi)$ from
\beq
 f'(\phi ) = \frac{\sqrt{3} M \,\phi}{\left(\phi +\sqrt{3}\right)}   \left(1-\frac{\phi ^2}{3}\right)^{(1-3 \sqrt{\alpha })/2}  \, ,
\eeq
where we have set $\langle T \rangle = 1/2$ to find $f (\phi)$. The solution to this 1st order equation has the form of a 
hypergeometric function
\begin{equation}
f (\phi) \; = \;   M \left[\frac{3-3^{-m} \left(3-\phi ^2\right)^{m+1}}{2 (m+1)}-\frac{\phi ^3 \, _2F_1\left(\frac{3}{2},-m;\frac{5}{2};\frac{\phi ^2}{3}\right)}{3 \sqrt{3}} \right]  \, ,
\label{alphastaro}
\end{equation}
where $m = \frac{3}{2} \left(\sqrt{\alpha} - 1 \right)$. 
Remarkably, when this expression for $f(\phi)$ is used in (\ref{Wf}), we obtain the following 
scalar potential 
\begin{equation}
V = \frac{3 M^2 \phi ^2}{\left(\phi +\sqrt{3}\right)^2},
\end{equation}
which, in terms of canonically-normalized fields~(\ref{kinphi}), yields the $\alpha$-Starobinsky model of inflation:
\begin{equation}
V = \frac{3}{4} M^2 \left(1 - e^{- \sqrt{\frac{2}{3 \alpha}} x} \right)^2.
\label{alstaro}
\end{equation}
It is important to note that the hypergeometric superpotential~(\ref{alphastaro}) is a function of a matter field $\phi$ only, 
therefore it can be successfully combined with the dark energy sector $W_{dS}$.

Despite its rather cumbersome form, the expression in (\ref{alphastaro}) simplifies dramatically for certain
values of $\alpha$. For example, for $\alpha = 1$, the superpotential is simply our original 
Wess-Zumino superpotential given in Eq. (\ref{wi})  as, in this case, $m=0$ and $_2F_1\left(\frac{3}{2},0;\frac{5}{2};\frac{\phi ^2}{3}\right)=1$. 
A relatively simple form for $f(\phi)$ also arises for $\alpha = 25/9$:
\beq
f = M \left( \frac{\phi ^5}{15 \sqrt{3}}-\frac{\phi ^4}{12}-\frac{\phi ^3}{3 \sqrt{3}}+\frac{\phi ^2}{2} \right) \, ,
\eeq
Other polynomial forms arise when $\alpha = 49/9$ and 9.
However, for any $\alpha$, the scalar potential always reduces to the $\alpha$-Starobinsky potential (\ref{alstaro}). 
We note that the full superpotential, $W_I$ is a polynomial whenever $9 \alpha$ is an odd perfect square other than 1.

\begin{figure}[h!]
\centering
\includegraphics[scale=0.7]{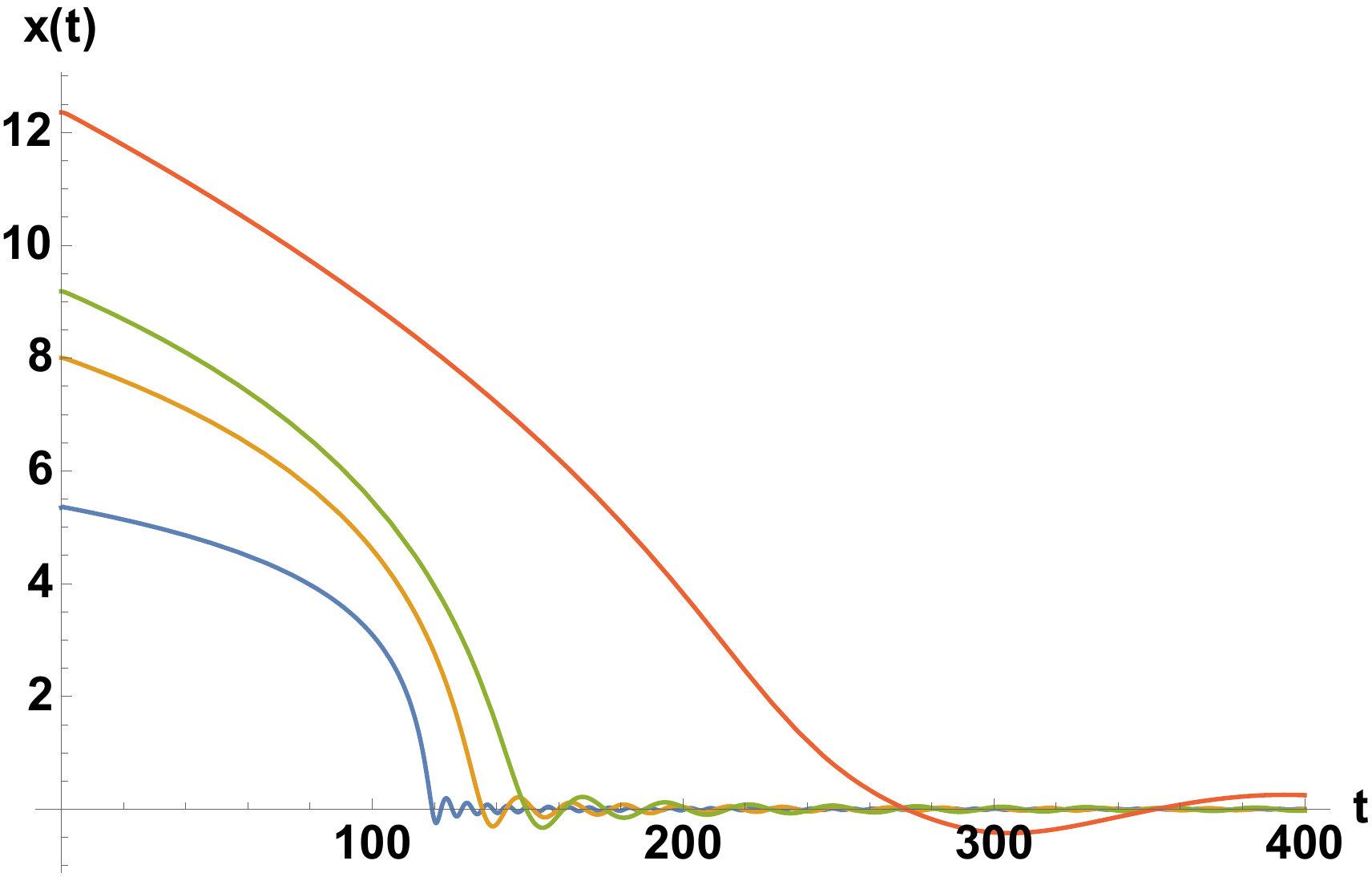}
\caption{\it Evolution of the inflaton field $x(t)$ in $\alpha$-Starobinsky models. The blue line shows the original Starobinsky inflationary potential with $\alpha = 1$, the yellow line shows 
the case with $\alpha = 5$, the green line shows the case with $\alpha = 10$, and the orange line shows the case with $\alpha = 30$. The units of time are $10^4/M_P$.}
\label{fig:astaro}
\end{figure}

Because the potential (\ref{alstaro}) depends on $\alpha$, the evolution of inflaton field and resulting slow-roll parameters differ when $\alpha$ is varied. We show in Fig.~\ref{fig:astaro} the slow-roll evolution of the field $x$ for different values of $\alpha$. Once again, we assume that all models have strongly stabilized moduli, so that we can safely treat them as single-field models of inflation. We consider the following $\alpha$-Starobinsky cases with four different values of $\alpha$ that give $N_* = 55$:
 \begin{tabular}{llll}
 \\
 \tabitem $\alpha = 1,$ & $\quad x(0) = 5.347,$ &$\quad r = 0.0035,$ & $\quad n_s = 0.965$. \\
 \tabitem $\alpha = 5,$ & $\quad x(0) = 8.003,$ &$\quad r = 0.0138,$ & $\quad n_s = 0.966$. \\
 \tabitem $\alpha = 10,$ & $\quad x(0) = 9.181,$ &$\quad r = 0.0230,$ & $\quad n_s = 0.967$. \\
 \tabitem $\alpha = 30,$ & $\quad x(0) = 12.354,$ &$\quad r =  0.0430,$ & $\quad n_s = 0.967$. \\
 \end{tabular} \\
 ~~\\
As could be expected from the form of the scalar potential, 55 e-folds of inflation can be obtained using increasing initial field
value,s $x(0)$ as $\alpha$ is increased. While $n_s$ varies little as $\alpha$ increases, the
tensor-to-scalar ratio, $r$, increases from its nominal Starobinsky inflation value of $r = 0.0035$
to $r = 0.0430$ when $\alpha = 30$. This effect is clearly seen in Fig. \ref{staroplot2}.

\section{Unified $STU$ Models}
In this Section we apply our framework to a more general effective supergravity theory that may emerge from string theory compactifications in the low-energy limit. We follow the treatment presented in~\cite{FK} and illustrate how to construct unified $STU$ inflationary models with supersymmetry breaking and a positive cosmological constant. In Section 3.2 we introduced multi-field no-scale attractors and argued that inflation must be driven by matter-like fields $\phi^i$ while moduli fields $T^n$ are responsible for supersymmetry breaking through $F$-terms. This Section shows how to relate the unified no-scale formulation to specific M/string theory models.

We begin by considering a seven-disk manifold, which can be connected to M-theory compactified on a 7-manifold with $G_2$ holonomy. We introduce the following K\"ahler potential:
\begin{equation}
K = -\sum_{i = 1}^7 \ln \left(T_i + T^{\dagger}	_i \right),
\label{seven}
\end{equation}
which describes a seven-disk manifold with $\left[\frac{SU(1,1)}{U(1)} \right]^7$ coset symmetry. Within the framework~(\ref{seven})
we consider the $STU$ model, which is described by the K\"ahler potential~(\ref{STU}). This expression arises in type IIB string theory and is characterized by a $\frac{T^6}{\mathbb{Z}_2 \, \cross \, \mathbb{Z}_2}$ coset manifold. The three complex scalar fields in~(\ref{STU}) are interpreted as a volume modulus $T$,
the axiodilaton $S$,  and a complex structure modulus $U$. 

We consider a specific unified $STU$ example, where we introduce an untwisted matter-like field $\phi$ into the K\"ahler potential 
$-3 \ln \left(T + T^{\dagger} \right)$. Thus, our model is characterized by the following $STU$ K\"ahler potential:
\begin{eqnarray}
K & = & -3 \ln \left[T + T^{\dagger} + \beta^R_{T} \left(T + T^{\dagger} - 1 \right)^4 + \beta^I_{T} \left(T - T^{\dagger} \right)^4 - \frac{|\phi|^2}{3} \right]  \nonumber \\
&& -3 \ln \left[U+ U^{\dagger} + \beta^R_{U} \left(U + U^{\dagger} - 1 \right)^4 + \beta^I_{U} \left(U - U^{\dagger} \right)^4 \right]  \nonumber \\
&& - \ln \left[S + S^{\dagger} + \beta^R_S \left(S + S^{\dagger} - 1 \right)^4 + \beta^I_S \left(S - S^{\dagger} \right)^4  \right],
\label{kah5}
\end{eqnarray}
where the matter-like field $\phi$ is identified as the inflaton field, and we have introduced quartic terms to stabilize the imaginary parts of $T, S$ and $U$. It should be noted that one can consider various $STU$ models, and the untwisted matter-like field $\phi$ can be also incorporated with axio-dilatonic K\"ahler potential $- \ln \left(S + S^{\dagger} \right)$ or complex structure modulus K\"ahler potential $- 3 \ln \left(U + U^{\dagger} \right)$. We follow the treatment introduced in Section 3.2 and consider the unified superpotential~(\ref{unisup1}), given by the combination of the Wess-Zumino superpotential $W_I$ and the supersymmetry breaking/dark energy sector superpotential $W_{dS}$:
\begin{eqnarray}
W & = & W_{I} + W_{dS} = M \left( \frac{\phi^2}{2} - \frac{\phi^3}{3 \sqrt{3}} \right) +  \nonumber \\
&& {\tilde \lambda}_1 M^3 \cdot  \xi_1^{\frac{3}{2} \left(1 - r_1 \right)} \xi_2^{\frac{3}{2} \left(1 - r_2 \right)} \xi_3^{\frac{3}{2} \left(\frac{1}{3} - r_3 \sqrt{\frac{1}{3}}\right)} - {\tilde \lambda}_2 M^3\cdot  \xi_1^{\frac{3}{2} \left(1 + r_1 \right)} \xi_2^{\frac{3}{2} \left(1 + r_2 \right)} \xi_3^{\frac{3}{2} \left(\frac{1}{3} + r_3 \sqrt{\frac{1}{3}}\right)},
\label{uniSTU}
\end{eqnarray}
where $\xi_1 = 2T - \frac{\phi^2}{3}$, $\xi_2 = 2U$ and $\xi_3 = 2S$. At the minimum, the strongly stabilized fields $T$, $U$, and $S$ acquire the vacuum expectation values $\langle T \rangle = \langle U \rangle = \langle S \rangle = \frac{1}{2}$, while the matter-like field has $\langle \phi \rangle = 0$, and the positive cosmological constant is given by $\Lambda = 12 {\tilde \lambda}_1 {\tilde \lambda}_2 M^6$.

In order to consider more complicated unified $STU$ models and connect them to M/string theory, we rewrite the general multi-field K\"ahler potential form~(\ref{kah4}) in $STU$ form:
\begin{eqnarray}
K & = & - \sum_{n=1}^3 \ln \left[T^n + T_n^{\dagger} + \beta^R_{T_n} \left(T^n + T_n^{\dagger} - 1 \right)^4 + \beta^I_{T_n} \left(T^n - T_n^{\dagger} \right)^4 - \frac{|\phi_T^n|^2}{3} \right]  \nonumber \\
&& -\sum_{a=1}^3 \ln \left[U^a + U_a^{\dagger} + \beta^R_{U_a} \left(U_a + U_a^{\dagger} - 1 \right)^4 + \beta^I_{U_a} \left(U^a + U_a^{\dagger} \right)^4 - \frac{|\phi^a_U|^2}{3} \right]  \nonumber \\
&& - \ln \left[S + S^{\dagger} + \beta_S^R \left(S + S^{\dagger} - 1 \right)^4 + \beta_S^I \left(S - S^{\dagger} \right)^4 - \frac{|\phi_S|^2}{3} \right],
\label{kah6}
\end{eqnarray}
where, in addition to the quartic stabilization terms in both real and imaginary directions for the complex fields $T^n$, $U^a$ and $S$,
we have also introduced three classes of matter-like fields $\phi_T^n$, $\phi_U^a$ and $\phi_S$, which are associated with prospective inflaton fields. It is important to note that one can always discard irrelevant matter-like fields in the K\"ahler potential~(\ref{kah6}) and consider various $STU$ model combinations, but for clarity we have provided the complete form. The quartic 
stabilization terms in K\"ahler potential~(\ref{kah6}), ensure that the moduli acquire vacuum expectation values $\langle T^n \rangle = \langle U^a \rangle = \langle S \rangle = \frac{1}{2}$.

\begin{figure}[ht!]
\centering
\includegraphics[scale=0.9]{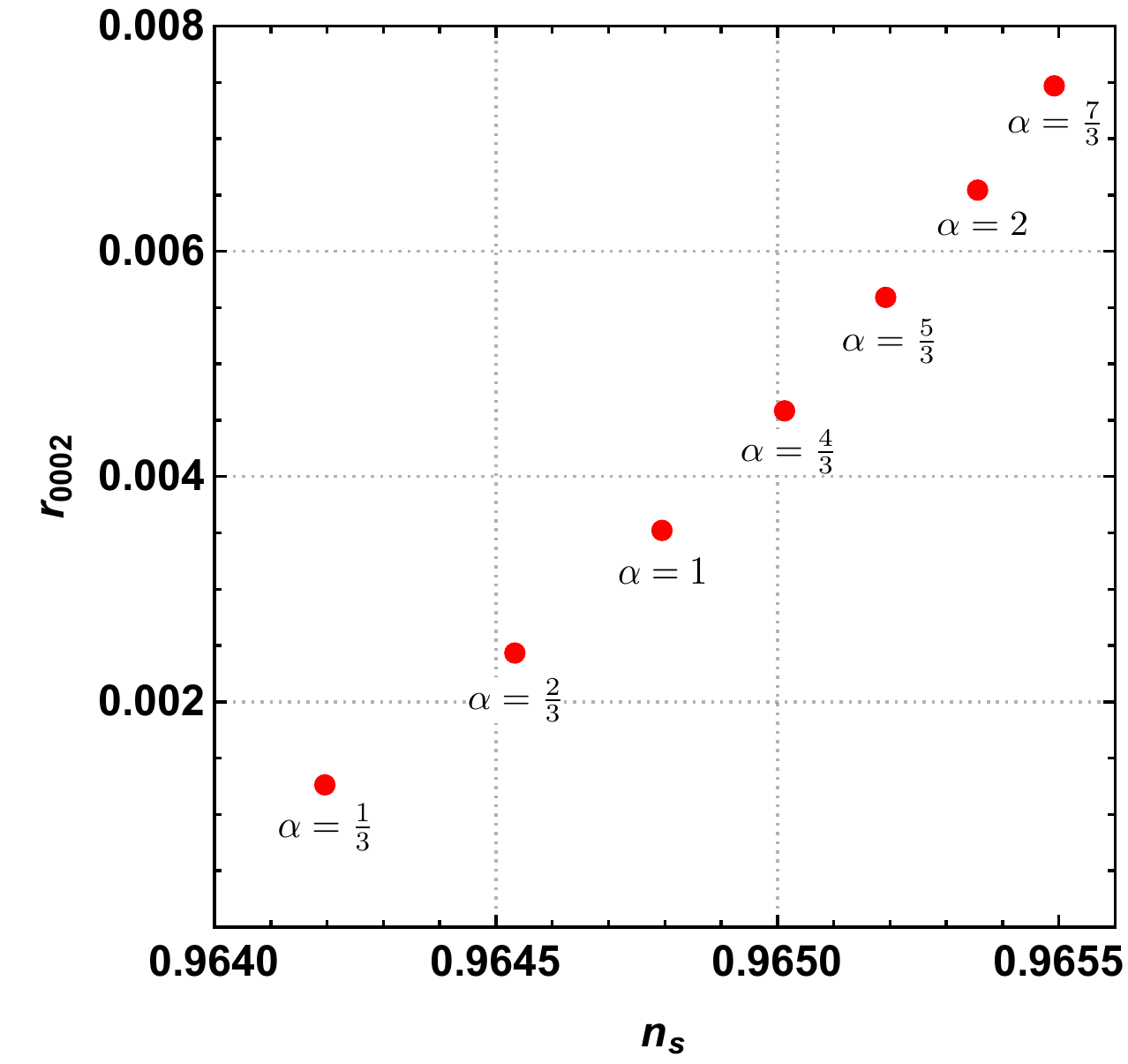}
\caption{\it Values of $n_s$ and $r$ for the choices of $\alpha$ in the $STU$ model.}
\label{staroplotstu}
\end{figure}

In this scenario, inflation can be driven by up to 7 different matter-like fields, and the principles discussed in Section 3.2 can be applied. As in \cite{FK}, we consider a specific example where we impose the following conditions on the form~(\ref{kah5}) of the K\"ahler potential:
\begin{equation}
T^n = U^a = S = T, \quad~\text{and} \quad~\phi_T^n = \phi_U^a = \phi_S = \phi,
\end{equation} 
and for clarity we exclude quartic stabilization terms. Comparing with the generic K\"ahler potential that parametrizes an $\frac{SU(2, 1)}{SU(2) \cross U(1)}$ coset manifold~(\ref{3}), we have
\begin{equation}
K \; = \; -3 \, \alpha \, \ln \left(T + T^\dagger - \frac{\phi \phi^\dagger}{3} \right) \; = \; -7 \,  \ln \left(T + T^\dagger - \frac{\phi \phi^\dagger}{3} \right) \, ,
\label{disk}
\end{equation}
i.e., we have the relation $3 \, \alpha = 7$. Alternatively, if we set a subset of the complex scalar fields $T^n$, $U^a$ and $S$ equal to $T$ and the associated matter-like fields $\phi_T^n$, $\phi_U^a$ and $\phi_s$ equal to $\phi$, we can remove the remaining trivial complex fields from the Lagrangian by setting them equal to a constant, and obtain the following curvature parameter values~\footnote{For a more detailed discussion on seven-disk manifold and no-scale geometry, see~\cite{FK}.}:
\begin{equation}
3 \alpha = \left \{ 1, 2, 3, 4, 5, 6, 7 \right \}. 
\label{values}
\end{equation}
Previously,  we have argued that for unified no-scale models inflation must be driven by a matter-like field $\phi$. Therefore, if we consider the $\alpha$-Starobinsky inflationary model, which is characterized by K\"ahler potential~(\ref{3}) and superpotential~(\ref{alphastaro}), with curvature parameter values~(\ref{values}) and a nominal choice of e-foldings $N_* = 55$, we obtain	 the following cosmological parameters:

 \begin{tabular}{llll}
 \tabitem $ \alpha = \frac{1}{3},$ & $\quad x(0) = 3.834$ &$\quad r = 0.001259,$ & $\quad n_s = 0.964$. \\
 \tabitem $\alpha = \frac{2}{3},$ & $\quad x(0) = 4.751,$ &$\quad r = 0.002430,$ & $\quad n_s = 0.965$. \\
 \tabitem $\alpha = 1,$ & $\quad x(0) = 5.347,$ &$\quad r =  0.003533,$ & $\quad n_s = 0.965$. \\
  \tabitem $\alpha = \frac{4}{3},$ & $\quad x(0) = 5.793,$ &$\quad r =  0.004581,$ & $\quad n_s = 0.965$. \\
 \tabitem $\alpha = \frac{5}{3},$ & $\quad x(0) = 6.150,$ &$\quad r =  0.005581,$ & $\quad n_s = 0.965$. \\
      \tabitem $\alpha = 2,$ & $\quad x(0) = 6.448,$ &$\quad r =  0.006539,$ & $\quad n_s = 0.965$. \\
         \tabitem $\alpha = \frac{7}{3},$ & $\quad x(0) = 6.705, $ &$\quad r =  0.007458,$ & $\quad n_s = 0.965$. \\
 \end{tabular}\\
 ~~\\
These cosmological parameter values are illustrated in Fig.~\ref{staroplotstu}. They are all comfortably consistent with the current
observational constraints shown in Fig.~2.

\section{Summary}

Intriguingly, observations of the CMB are highly consistent with the original Starobinsky model of inflation, whereas many other
models proposed subsequently have fallen by the wayside. The range of $n_s$ preferred by the data is highly consistent with
the Starobinsky prediction $n_s = 1 - 2/N_*$, where $N_* \sim 50$ to 60 is the number of e-folds of inflation, and the current
upper limit on $r \sim 0.06$ is also consistent with the Starobinsky prediction $ \sim 0.003$, albeit with considerable leeway.
One of our reasons for being intrigued by the Starobinsky model is that its predictions are shared by simple models based on
no-scale supergravity (\ref{3}), as was first discussed in~\cite{ENO6} for the case $\alpha = 1$. As we have emphasized there and in this paper, 
some form of no-scale supergravity emerges naturally as the effective low-energy theory derived in compactified string models, 
thus offering a specific bridge between cosmological observables and string theory.

As was first emphasized in~\cite{Avatars}, one may consider generalizations of the original model (\ref{3}) with $\alpha \ne 1$,
depending on the (combination of) compactification modulus field(s) providing the inflaton, yielding the predictions (\ref{avatars})
for $n_s$ and $r$. Similar models were discussed from a more general point of view in~\cite{alpha1,alpha2,att2,rs,att3,att4,otheratt,FK}, where they were dubbed
$\alpha$-attractors. The predictions of such models are compatible with the CMB data for a large range of possible values of $\alpha$,
as seen in Fig.~\ref{staroplot2}. In this connection, we are encouraged by the recent approval of the {\tt LiteBIRD} space mission,
which is projected to be able to measure $r$ to an accuracy of $\pm 0.001$, sufficient to measure $\alpha$ with interesting
precision, and thereby provide an entr\'ee into the phenomenology of string compactification.

The main purpose of this paper has been to develop a framework for this prospective phenomenology that extends beyond the scope of
K\"ahler manifolds with the $\frac{SU(2,1)}{SU(2) \cross U(1)}$ coset structure considered previously~\cite{ENOV1,ENOV2}.
In~\cite{ENOV1} we gave a general classification of superpotentials for such manifolds that lead to Starobinsky-like inflation when $\alpha = 1$,
discussing the relations between them provided by the underlying no-scale structure. Then, in~\cite{ENOV2} we
showed how such Starobinsky-like predictions for the CMB observables could be combined in a unified framework with modulus fixing, 
supersymmetry breaking and a small cosmological constant compatible with the current density of dark energy.
In this paper we have extended these earlier constructions in two main directions: to inflationary models based on generalized no-scale structures 
with  $\frac{SU(N,1)}{SU(N) \cross U(1)}$ coset structures and to models with different values of $\alpha$ and hence $r$, as may occur if
the inflaton corresponds to only a subset of the complex K\"ahler moduli, or if complex structure moduli also help drive inflation.
As in the previous  $\frac{SU(2,1)}{SU(2) \cross U(1)}$ case, key building blocks in these generalizations are played by minimal
superpotentials that yield Minkowski vacua and can, in pairs, yield either de Sitter or anti de Sitter vacuum states.

We plan to return in a forthcoming paper to more detailed phenomenological investigations of such generalized $\alpha$-no-scale models,
with a view to kickstarting the exploration of cosmological string phenomenology that will be opened up by {\tt LiteBIRD} and other CMB
experiments. We have emphasized the important role that could be played by measurements of $r$ in constraining the geometry of
the the underlying no-scale K\"ahler manifold but, before closing, we stress also the importance of measurements of the scalar tilt, $n_s$.
As seen in (\ref{3}), this is sensitive to the number of e-folds of inflation, $N_*$, and hence to the post-inflationary history of the Universe.
In particular, it is sensitive to the amount of post-inflationary reheating, and hence to the coupling of the inflaton to lighter degrees of
freedom. Thus, it complements the measurement of $r$ by being sensitive to the superpotential of the inflationary model. Combining
the constraints on $r$ and $n_s$ could provide unique insights into the underlying string compactification.

\subsection*{Acknowledgements}

\noindent
The work of JE was supported in part by the United Kingdom STFC Grant
ST/P000258/1, and in part by the Estonian Research Council via a
Mobilitas Pluss grant. The work of DVN was supported in part by the DOE
grant DE-FG02-13ER42020 and in part by the Alexander~S.~Onassis Public
Benefit Foundation. The work of KAO was
supported in part by DOE grant DE-SC0011842 at the University of
Minnesota.

\end{document}